\documentclass[aps,floatfix,prd,twocolumn,a4paper,10pt,citeautoscript,
preprintnumbers,superscriptaddress,showpacs,longbibliography]{revtex4-1} 

\usepackage[english]{babel} 	
\usepackage{amsmath}        	
\usepackage{amssymb}        	
\usepackage{amsfonts}
\usepackage{bm}					
\usepackage{bbm}            	
\usepackage{empheq}
\usepackage[normalem]{ulem}

\usepackage{graphicx}       
\usepackage{graphics}       
\usepackage{color}
\DeclareGraphicsExtensions{.jpg,.pdf} 
\usepackage{hyperref}
\hypersetup{colorlinks=true, pdfstartview=FitV, 
linkcolor=blue, citecolor=blue, urlcolor=blue}

\renewcommand{\theequation}{\arabic{equation}}
\newcommand{\Equation}[2]{\begin{equation}\label{#1}#2\end{equation}}

\newcommand{\bs}{\boldsymbol}
\newcommand{\Figref}[1]{Fig.~\ref{#1}}
\newcommand{\Eqref}[1]{\eqref{#1}}
\newcommand{\Real}{\mathrm{R}}  					
\newcommand{\groupCP}[1]{\mathrm{C}\mathbb{P}^{#1}} 
\newcommand{\Exp}[1]{\text{e}^{#1}}

\newcommand{\Curl}{{\bs\nabla}\times}

\newcommand{\Q}{{\cal Q}}

\newcommand{\bsigma}{{\boldsymbol \sigma}}
\newcommand{\bnabla}{{\boldsymbol \nabla}}
\newcommand{\btau}{{\boldsymbol \tau}}

\begin{document}
\graphicspath{{../Final-plots/}}
\title{Nematic Skyrmions in Odd-Parity Superconductors}

\author{A.~A.~Zyuzin}
\affiliation{Department of Physics, KTH-Royal Institute of Technology, Stockholm, SE-10691 Sweden}
\affiliation{Ioffe Physical--Technical Institute,~194021 St.~Petersburg, Russia}

\author{Julien~Garaud}
\affiliation{Department of Physics, KTH-Royal Institute of Technology, Stockholm, SE-10691 Sweden}

\author{Egor~Babaev}
\affiliation{Department of Physics, KTH-Royal Institute of Technology, Stockholm, SE-10691 Sweden}

\begin{abstract}

We study topological excitations in two-component nematic superconductors, 
with a particular focus on $\mathrm{Cu_xBi_2Se_3}$ as a candidate material. 
We find that the lowest-energy topological excitations are coreless vortices: 
a bound state of two spatially separated half-quantum vortices. These objects 
are nematic Skyrmions, since they are characterized by an additional
topological charge.
The inter-Skyrmion forces are dipolar in this model, i.e. attractive for 
certain relative orientations of the Skyrmions, hence forming multi-Skyrmion 
bound states.
\end{abstract}
\maketitle

Bulk superconductivity in topological insulator materials has recently 
been observed in electron-doped $\mathrm{Cu_xBi_2 Se_3}$ in Refs.~\cite{Hor2010,
Wray_CuBiSe,Kriener_CuBiSe} and unusual superconducting states in this system 
were theoretically considered in Ref.~\cite{FuBerg2010}. There, it was argued 
that the fully-gapped single order parameter superconductor, which has a 
spin-triplet pairing with odd parity and that possesses topologically protected 
gapless surface states, is favored over other superconducting states.
It was later put forward \cite{Fu_nematic} that the interplay of the crystal 
lattice anisotropy and the nematic superconductivity might be consistent with 
Knight-shift anisotropy measurements in $\mathrm{Cu_xBi_2 Se_3}$, that show 
spontaneous breaking of the spin-rotation symmetry below the superconducting 
transition temperature \cite{Matano.Kriener.ea:16}.  
In this model, the nematic superconducting state has an odd-parity spin-triplet 
pairing and is described by a two-component order parameter that spontaneously 
breaks the rotational symmetry in the basal plane of the lattice. This scenario 
of nematic superconductivity is supported by the recent observation of two fold 
rotational symmetry of the magnetic field in specific heat and upper critical 
field measurements of the superconducting state \cite{Yonezawa_specificheat}. 
Bulk superconductivity was also reported in $\mathrm{Sr_xBi_2Se_3}$ 
\cite{Shruti_SrBiSe,Liu_SrBiSe,Pan_Hc2} and in magnetically doped 
$\mathrm{Nb_xBi_2Se_3}$ \cite{Qui_NbBiSe, Asaba_NbBiSe}, where upper critical 
field \cite{Pan_Hc2} and magnetic torque measurements \cite{Asaba_NbBiSe} 
reveal signatures of rotational symmetry breaking in the amplitude of the 
superconducting gap.

Nematic superconducting states have interesting properties that were recently 
theoretically addressed. For example, such states show a specific anisotropy 
of the upper critical magnetic field \cite{Vanderbos_Hc2}, undergo phase 
transitions to superconducting states that break the time-reversal symmetry 
as a result of the interplay of ferromagnetism and superconductivity 
\cite{Yuan_Ferro, Guinea_Ferro}, and host Majorana fermions at the surface 
\cite{Tanaka_Majorana,Venderbos.Kozii.ea:16a}. This raises the question of 
the properties of topological excitations in this kind of materials. 
Recent work \cite{Martin_Majorana} presented an ansatz-based investigation of 
Kramers pairs of Majorana fermions bound inside a specific type of composite 
vortices that do not carry magnetic flux. In the model that we consider below, 
such an ansatz describes unstable vortex solutions.

In this Letter we address the question of the nature of the lowest-energy 
vortex excitations in nematic superconductors. To this end, we investigate 
vortex solutions in a two-component Ginzburg-Landau (GL) model consistently 
derived from the microscopic theory. We find that the lowest-energy topological 
excitations are coreless and consist of two spatially separated half-quantum 
vortices (HQVs) \footnote{Half-quantum vortices are of substantial current 
interest in various contexts due to proposals that they host Majorana fermions, 
see for example Refs.~\cite{Read.Green:00,Ivanov:01}.} 
\nocite{Read.Green:00,Ivanov:01}, such that the total superconducting density 
has no zeros. These excitations can be characterized by a Skyrmion topological 
index, which is zero for singular vortices. 
Heuristically, the Skyrmion terminology follows from the fact that the coreless 
vortex can be seen as a texture of a unit vector that fully covers the target 
two-sphere. The unit vector that maps to the target two-sphere is defined as a 
projection of the superconducting degrees of freedom onto the vector composed 
of the Pauli matrix set.

Such a coreless vortex, which is a bound state of two HQVs, shows as a dipolelike 
configuration of the relative phase between the components of the order parameter, 
and thus can mediate a long-range dipole interaction between the Skyrmions, 
thus binding them together into a multi-Skyrmion bound state. 

We consider a model of a three-dimensional topological insulator in the 
presence of a magnetic field, having in mind $\mathrm{Bi_2Se_3}$ as a 
particular material candidate, which is a narrow gap semiconductor with 
a layered crystal structure. The system is described by the Hamiltonian 
$\mathcal{H}=\int \Psi^{\dag}(\mathbf{r})H(\mathbf{r})\Psi(\mathbf{r}) d^3r$, 
with
\begin{eqnarray}\label{Hamiltonian_Main}\nonumber
H(\mathbf{r})&=&
v\tau_z\left[\bsigma \times\bigg(-i\bnabla-\frac{e}{c}\mathbf{A}(\mathbf{r})\bigg)\right]\cdot \hat{z}\\
&+&v_z\tau_y\bigg(-i\nabla_z-\frac{e}{c}A_z(\mathbf{r})\bigg)+m\tau_x,
\end{eqnarray}
where $\mathbf{A}(\mathbf{r})$ is the vector potential, $e<0$ is the electric 
charge, and $m$ describes the coupling between the orbitals of $\mathrm{Bi_2Se_3}$.
Here, $v$ and $v_z$ are the Fermi velocities that characterize the anisotropic 
dispersion of the massive Dirac fermion in the absence of the magnetic field: 
$E_{\pm}(\mathbf{p})=\pm\{v^2(p_{x}^2+p_y^2)+v_z^2p_z^2+m^2\}^{1/2},$
where $\mathbf{p}=(p_x, p_y, p_z)$ is the momentum of a particle. 
The Pauli matrices $\sigma_a$ and $\tau_a$ (with $a=x,y,z$), respectively, describe 
the real spin ($\uparrow,\downarrow$) and the orbital pseudospin ($1,2$) degrees 
of freedom.  
The electron operator is given by $\Psi(\mathbf{r})=(\Psi_{\uparrow,1}(\mathbf{r}),
\Psi_{\downarrow,1}(\mathbf{r}),\Psi_{\uparrow,2}(\mathbf{r}),
\Psi_{\downarrow,2}(\mathbf{r}))^{\mathrm{T}}$, and $\hbar=1$ units are used 
here, and spin and pseudospin indices are omitted for clarity of notation throughout 
the Letter. The Zeeman contribution of the magnetic field to the Hamiltonian 
\Eqref{Hamiltonian_Main} is neglected compared to that the orbital effect. 
We also note that, although there is strong spin-orbit interaction in each orbital, 
the inversion symmetry of the system is preserved.
  
As demonstrated in Ref.~\cite{FuBerg2010} the electron-phonon interaction might 
lead to several distinct $s$-wave superconducting instabilities in this system: 
intraorbital spin-singlet, interorbital spin-singlet, and interorbital 
spin-triplet. Motivated by the experimental signatures for the nematic 
superconductivity, we focus here on the interorbital spin-triplet pairing, 
which is described by the interaction Hamiltonian within the 
Bardeen--Cooper--Schrieffer (BCS) approximation: 
\begin{eqnarray}
\mathcal{H}_{\mathrm{BCS}}=-\sum_{\sigma,\sigma'} \int d^3r
\bigg[ \Psi^{\dag}_{\sigma 1}(\mathbf{r})\Psi^{\dag}_{\sigma' 2}(\mathbf{r})
\Delta_{\sigma'\sigma}(\mathbf{r}) + \mathrm{h.c.}\bigg]\,, ~~
\end{eqnarray} 
where $\Delta_{\sigma'\sigma}(\mathbf{r}) = 
\lambda\langle \Psi_{\sigma' 2}(\mathbf{r})\Psi_{\sigma 1}(\mathbf{r})\rangle$ 
with the interaction constant $\lambda>0$. Note that we consider the zero 
harmonic of the electron-phonon interaction potential, which shall give a higher 
temperature of the superconductor-metal phase transition than higher harmonics. 
To proceed we introduce the Bogolyubov-de Gennes (BdG) Hamiltonian
$
\mathcal{H}_{\textrm{BdG}}=\frac{1}{2}\int \Phi^{\dag}(\mathbf{r})
H_{\textrm{BdG}}(\mathbf{r})\Phi(\mathbf{r}) d^3r
$, where
\begin{equation}
H_{\textrm{BdG}}(\mathbf{r}) = \bigg[
 \begin{matrix}
  H(\mathbf{r}) -\mu & \Delta(\mathbf{r}) \\
  \Delta^{+}(\mathbf{r}) & -\sigma_y H^{*}(\mathbf{r})\sigma_y+\mu
 \end{matrix}
\bigg]
\end{equation}
is written in the Nambu notations: 
$\Phi^{\dag}(\mathbf{r})=(\Psi^{\dag}(\mathbf{r}),
\Psi^{\textrm{T}}(\mathbf{r})(-i\sigma_y))$.
We assume the Fermi level to be in the conduction band and hence set $\mu>|m|$. 
In what follows we consider the interorbital spin-triplet pairing of the form 
$\Delta(\mathbf{r}) = \bsigma\cdot\boldsymbol{\Delta}(\mathbf{r})\tau_y$, where 
$\boldsymbol{\Delta}(\mathbf{r})=[\Delta_x(\mathbf{r}),\Delta_y(\mathbf{r}),0]$, 
such that $\Delta_x(\mathbf{r}) = -i[
\Delta_{\uparrow\uparrow}(\mathbf{r})-\Delta_{\downarrow\downarrow}(\mathbf{r})]/2$, 
and $\Delta_y(\mathbf{r}) = -[
\Delta_{\uparrow\uparrow}(\mathbf{r})+\Delta_{\downarrow\downarrow}(\mathbf{r})]/2$.
The strong spin-orbit interaction locks electron spin to the momentum, thus fixing 
the orientation of vector $\Delta$ to $[\Delta_x(\mathbf{r}),\Delta_y(\mathbf{r}),0]$.

In order to investigate the structure of the topological excitations, we derive 
microscopically the GL free energy functional for the two-component order 
parameter $\Delta_{\pm} = (\Delta_x \pm i \Delta_y)/\sqrt{2}$ (for details of 
the derivation, see Supplemental Material 
\cite{
[{The details of the derivation are given in the Supplemental Material (SM)}]
[{}] Supplementary}
). 
The scaled Ginzburg-Landau free energy functional reads as $4\pi F = h^2_0 l^3_m 
\int \left(\mathcal{F}(\mathbf{R})+ [\Curl\mathbf{a}(\mathbf{R})]^2\right) d^3R$, 
where
\begin{align}
&\mathcal{F}=\sum_{s=\pm}\bigg\{ -|\Delta_{s}|^2+ |D_x\Delta_{s}|^2
					+|D_y\Delta_{s}|^2+\beta_z |D_z\Delta_{s}|^2
	\nonumber \\
&+\beta_{\perp}(D_{-s}\Delta_{s})^{*} D_{s} \Delta_{-s}
+\frac{|\Delta_{s}|^4}{2}+ \frac{\gamma}{2}|\Delta_{s}|^2|\Delta_{-s}|^2
\bigg\}
\,,\label{GL}
\end{align}
where $\Delta_{\pm} = |\Delta_\pm|\Exp{i\varphi_\pm}$. Here we have defined 
the order parameter $\Delta_s(\mathbf{R})$ in the scaled form. The explicit 
scaling transformation for coordinate $\mathbf{R}$, vector potential 
$\mathbf{a}(\mathbf{R})$, and operator $D_{\pm}=D_x\pm iD_y$ in which 
$\mathbf{D}=-(i/\kappa)\boldsymbol{\nabla}_R+ \mathbf{a}(\mathbf{R})$ is given 
in Supplemental Material \cite{Supplementary}. In the dimensionless 
units, the coupling constant $\kappa$ parametrizes the magnetic field penetration 
length relative to a characteristic length scale associated with density variation. 
Its explicit expression through the parameters of the microscopic model is given 
in the Supplemental Material \cite{Supplementary} [\Eqref{Eq:kappa}], as well as the 
expression for other coefficients $h_0$, $l_m$, $\beta_{\perp,z}$, and $\gamma$. 
For $\kappa\ll 1$ the vortex core energy is relatively large and the 
thermodynamically stable Skyrmions do not form.

The anisotropy of the electronic spectrum results in the anisotropic gradient 
terms on the first line of the GL functional \Eqref{GL}. The interplay of the 
two-component order parameter and strong spin-orbit interaction gives 
rise to the mixed gradient term with the coefficient $0<\beta_{\perp}<1$ on 
the second line in Eq.~\Eqref{GL}. The GL free energy functional density is 
invariant under the joint rotation of coordinates and the components of the order 
parameter \footnote{For example, for a rotation of coordinates 
$(x,y)\rightarrow (y,-x)$ the components of the order parameter rotate 
$(\Delta_x,\Delta_y)\rightarrow- (\Delta_y,-\Delta_x)$.}.
Finally, the value of $\gamma$ determines whether the superconductor is in the 
nematic $0<\gamma<1$ or in the chiral $\gamma>1$ phase. Indeed, in the 
spatially homogeneous case, when $0<\gamma<1$ the free energy is minimal 
if the order parameter has the form $\Delta=\Delta_0(\cos\theta,\sin\theta,0)$ 
(for some real constant $\theta$ and $|\Delta_0|=1/\sqrt{1+\gamma}$), while 
it reads as $\Delta=\tilde{\Delta}_0(1,\pm i,0)$ (where $|\tilde{\Delta}_0|=1$) 
when $\gamma>1$.

Time-reversal symmetry is preserved in the nematic state. The chiral state, on 
the other hand is characterized by the nonzero electron spin polarization 
$\propto|\boldsymbol{\Delta}(\mathbf{r})\times\boldsymbol{\Delta}^{*}(\mathbf{r})|
\neq 0$. Interestingly, nontrivial pseudospin polarization, antiferromagnetic 
spin orientation in two orbitals, shows up in the first gradients of the order 
parameter in both chiral and nematic cases, see Supplemental Material 
\cite{Supplementary}.

The GL equation for the component $\Delta_s$ of the order parameter is 
obtained by the functional variation of Eq.~\Eqref{GL} with respect to 
$\Delta_s^*$: 
\begin{align}\nonumber
(D_x^2+D_y^2+\beta_z D_z^2)&\Delta_s +
\left( |\Delta_s|^2 +\gamma|\Delta_{-s}|^2-1\right)\Delta_s \\
&= -\beta_{\perp}D_{s}^2\Delta_{-s}\,,
\end{align} 
and is supplemented by the boundary condition
\begin{equation}
\mathbf{N}\cdot \mathbf{D}\Delta_s+(\beta_z-1)N_zD_z\Delta_s 
+N_sD_s\Delta_{-s}~\bigg|_{\mathrm{surf}}=0\,,
\end{equation}
where $N_s=N_x+isN_y$ and $\mathbf{N}$ is the unit vector directed normal to 
the surface. Note that here we do not consider the effects of the localized 
surface states. Finally, the vector potential $\mathbf{a}(\mathbf{R})$ of the 
magnetic field $\mathbf{b}(\mathbf{R})=\bnabla_R \times\mathbf{a}(\mathbf{R})$ 
satisfies 
\begin{align}\nonumber
&-2\Curl\mathbf{b}=\sum_{s=\pm}\bigg\{[\Delta_s^*\mathbf{D}\Delta_s+
(\beta_z-1)\Delta_s^* \hat{z} D_z\Delta_s+\mathrm{c.c.}]\\
&+(\hat{x}+is\hat{y})\beta_{\perp}[\Delta_s^*D_s\Delta_{-s}+\Delta_{-s}(D_{-s}\Delta_{s})^*]\bigg\}.
\end{align}

\begin{figure}[!htb]
\hbox to \linewidth{ \hss
\includegraphics[width=0.99\linewidth]{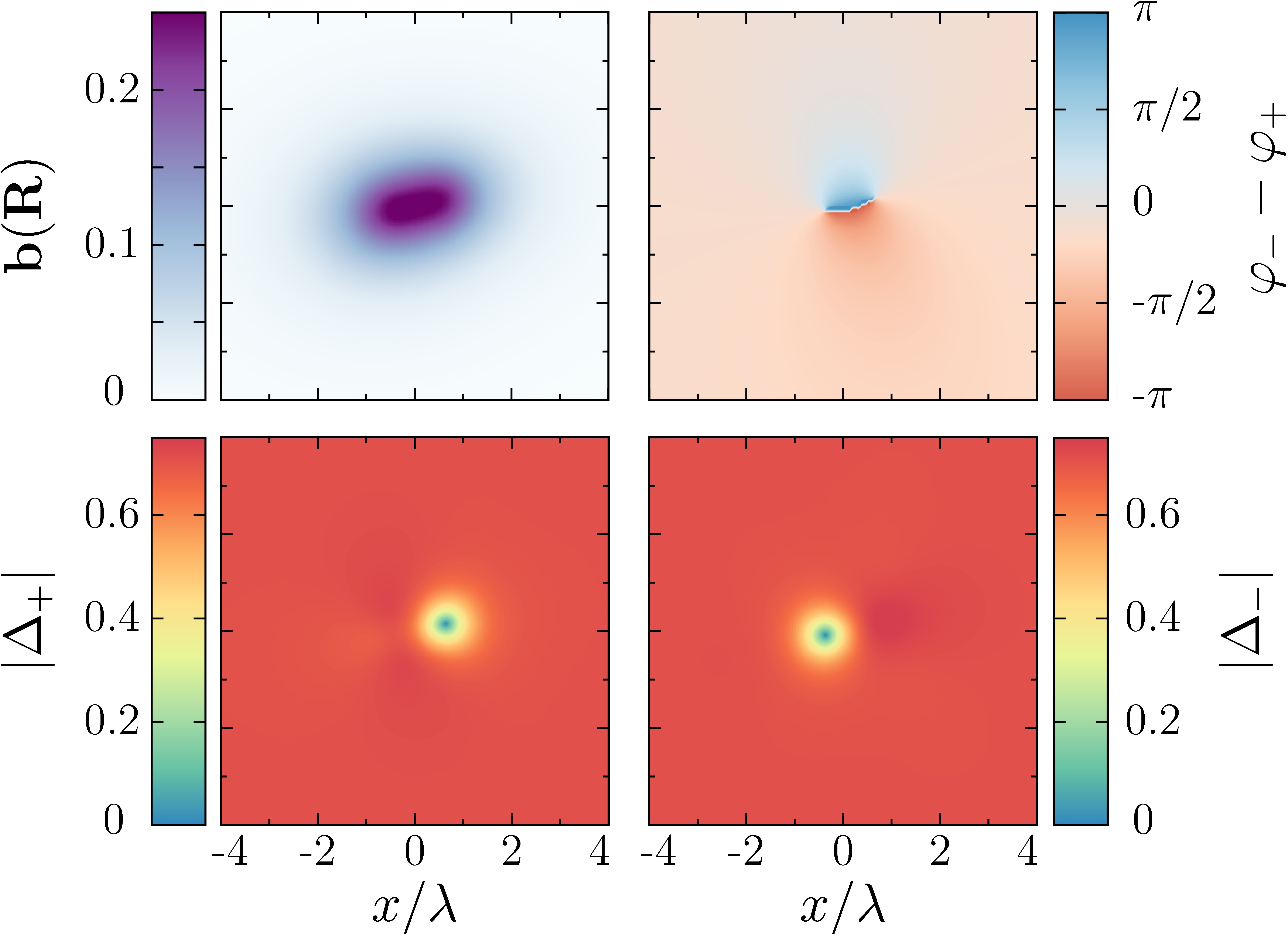}
\hss}
\caption{
%
Close-up view of the vortex core structure in the nematic phase in 
which we set $v=v_z$, $\beta_z=0$, and for $\kappa=3$.
On the first line, the first panel shows the distribution of the 
magnetic field, while the second displays the relative phase between 
the two components of the order parameter. The second line shows the 
densities of the two components of the order parameter. 
Clearly, the cores in both components do not superimpose, thus implying 
that vortices in the nematic phase are coreless defects.
Since cores do not overlap, the relative phase has $\pm2\pi$ winding 
around each core. Furthermore, the relative phase exhibits a dipolar mode 
that is long-ranged. 
}
\label{Fig:Nematic}
\end{figure}

We now turn to the investigation of the nature of topological excitations in 
the nematic superconductor. In two-component models, due to the coupling of 
the components to the vector potential $\mathbf{a}(\mathbf{R})$, the only 
solutions with finite energy per unit length have the same phase winding in both 
components of the order parameter, that is, a bound state of vortices in the
different components, each carrying a fraction of magnetic flux that adds up 
to a single flux quantum.
In the current model of the nematic superconductor vortices in each component 
of the order parameter carry half of a magnetic flux quantum; hence, they are 
half-quantum vortices (HQVs)
\footnote{This follows from that both components have the same ground state 
density. In general this 'symmetry' can be broken, leading to a more 
generic fractional quantization of the flux, see detailed discussion in 
\cite{frac}.}. \nocite{frac}
Typically, the magnetic interaction between HQVs favors cocentricity of 
the vortex cores in the different components (see, e.g., a detailed discussion 
in Ref.~\cite{frac}). On the other hand, the model considered here also features 
mixed gradients and biquadratic density-density terms that result in the 
repulsion between the cores of the half-quantum vortices. Provided the latter 
dominate, the competition between those forces may result in a bound state of 
nonoverlapping half-quantum vortices, thus breaking the axial symmetry of 
the solution. 

\begin{figure}[!htb]
\hbox to \linewidth{ \hss
\includegraphics[width=0.99\linewidth]{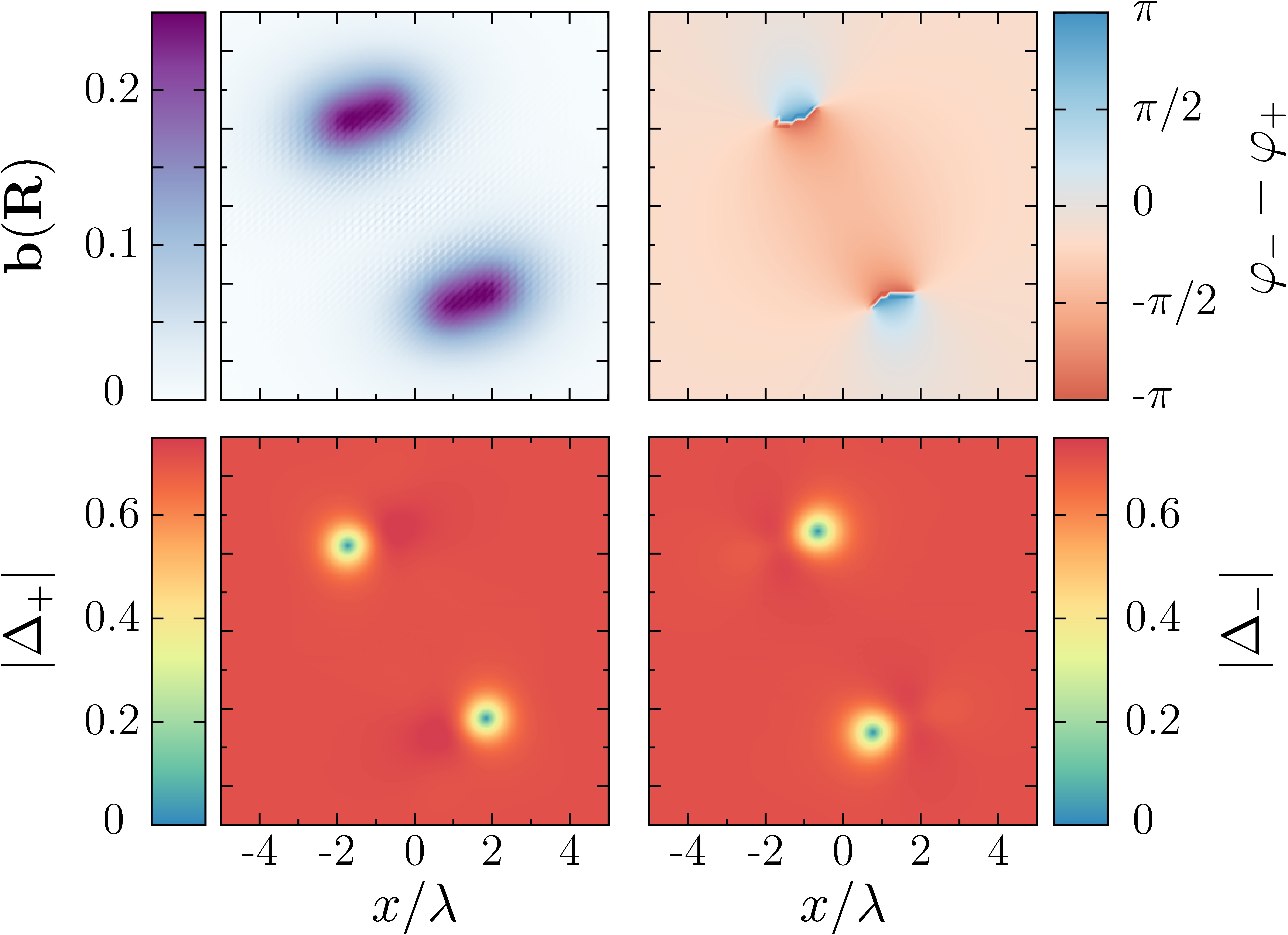}
\hss}
\caption{
%
A close-up view of a bound state of two coreless vortices carrying one 
flux quantum each. Each coreless vortex is a well-localized bound state 
of two HQVs. The dipolelike forces in the relative phase yields a long-range 
attraction that binds the single-quantum vortices together. Note that the 
dipoles are antialigned in the bound state as dictated by relative phase 
interaction. Displayed quantities are the same as in \Figref{Fig:Nematic}. 
} 
\label{Fig:Nematic2}
\end{figure}

To address whether vortices are singular (cocentered HQV) or coreless 
(i.e., noncocentered HQV), we numerically construct vortex solutions by 
minimizing the free energy \Eqref{GL}, starting by an initial configuration, 
in which both components $\Delta_\pm$ have the same winding. The theory is 
discretized within a finite-element formulation \cite{Hecht:12}, and minimized 
using a nonlinear conjugate gradient algorithm
\cite{
[{For detailed discussion on the numerical methods, see for example related 
discussion in: }]
[{}] Garaud.Babaev.ea:16}. 
Minimization procedure leads, after the convergence of the algorithm, to a vortex 
configuration that carries a number of flux quanta that is specified by the 
initial phase winding.
Figure \ref{Fig:Nematic} shows such a single-quantum vortex configuration in the 
model \Eqref{GL} for the nematic superconductor. Note that the picture shows a 
close-up view, displaying only a small part of the simulated numerical grid, which 
is chosen to be large enough so that vortices do not interact with the boundaries. 
Clearly, the vortex solution is not axially symmetric. Inspection of the core 
structure reveals that HQVs in different components are spatially separated 
and thus that this bound state of HQVs is coreless; i.e., there is no singularity 
of total density of superconducting components: 
$[|\Delta_{+}(\mathbf{R})|^2+|\Delta_{-}(\mathbf{R})|^2]^{1/2}$.
We simulated vortex solutions for various initial guesses that always converge 
to configurations as in \Figref{Fig:Nematic}. All investigated values of the 
parameter $\kappa$ led to coreless vortices in the type-II regime. The distance 
between HQVs is determined by the competition between magnetic attraction and 
repulsion mediated by other terms such as density-density interaction. This cannot 
be addressed analytically, but quantitatively it can be seen that increasing the 
value of $\kappa$ decreases binding the HQVs thus increasing their separation.
Since the superconductor is substantially away from the type-I regime, and because 
these excitations are energetically cheaper than singular vortices, a lattice of 
Skyrmions will form in external field.

\begin{figure*}[!htb]
\hbox to \linewidth{ \hss
\includegraphics[width=0.85\linewidth]{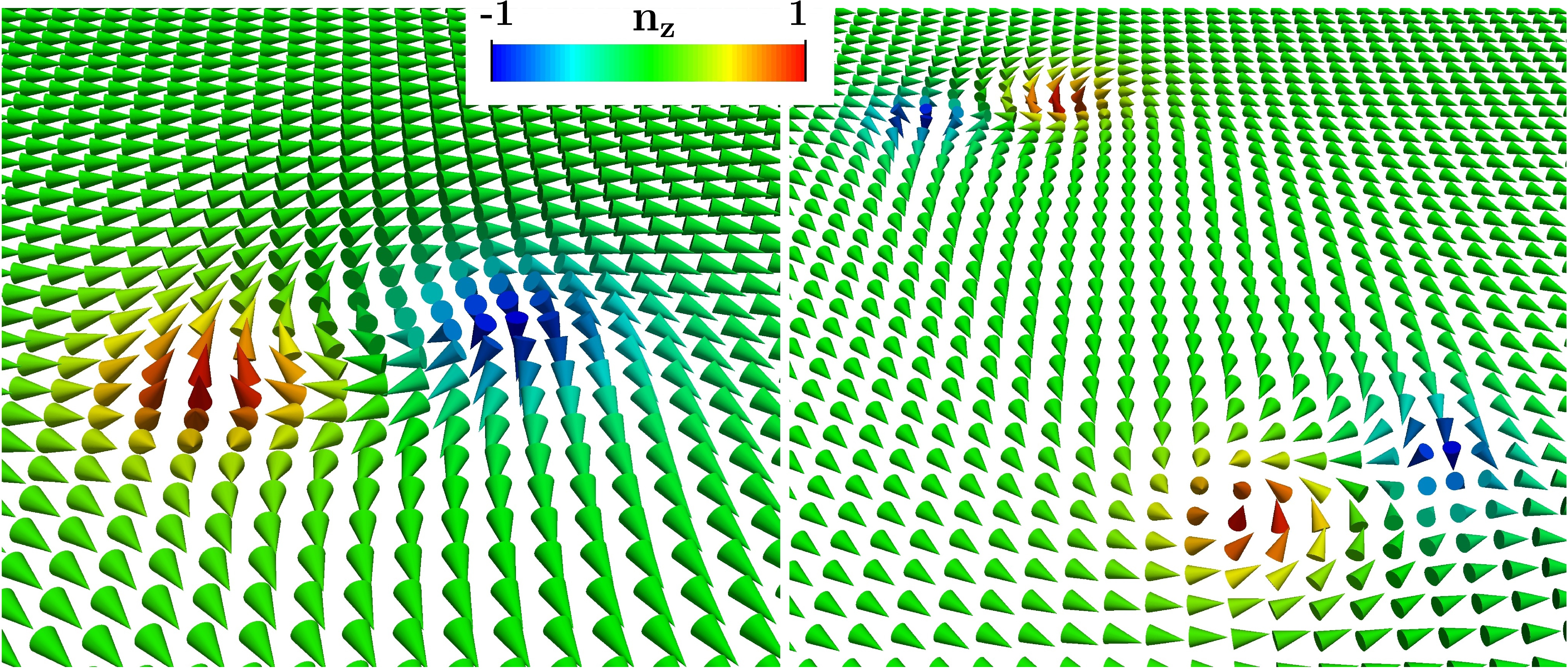}
\hss}
\caption{
%
Texture of the unit vector ${\bf n}$ defined as the projection of the 
superconducting degrees of freedom onto the spin-$1/2$ Pauli matrices. 
The left panel shows a single-quanta solution corresponding to \Figref{Fig:Nematic}, 
while the right panel corresponds to the bound state displayed in 
\Figref{Fig:Nematic2}.
Inspection of the single-quanta solution show how ${\bf n}$ wraps the target 
two-sphere. The north (respectively, south) pole signals zero of $\Delta_+$ 
(respectively $\Delta_-$), and thus the position of respective HQVs.
From this one, can see structural difference of nematic Skyrmions, compared to 
Skyrmions in chiral superconductors \cite{Garaud.Babaev:15a}. Here the solution 
has clearly the form of weakly interacting well-separated Skyrmions with unit 
topological charge each. By contrast, in chiral case the Skyrmionic topological 
charge tends to be relatively uniformly spread along closed domain walls.
} 
\label{Fig:Texture}
\end{figure*}
%
 
As singularities in both components do not overlap, there is a dipolelike 
configuration of the relative phase $\varphi_--\varphi_+$ between the 
components. Importantly,  the phase-difference gradients are very strong.  
This indicates that intervortex forces include torque and a long-range dipole 
interaction, that can lead to a long-range attraction between single-quanta 
vortices. As displayed in \Figref{Fig:Nematic2}, by initially setting a double 
phase winding in each component, we find that indeed two single-quantum 
vortices form a bound state due to the dipolar forces.
Note that the presence of dipolar interactions usually has clear signatures 
in structure formation. In particular, studies of Skyrmion solutions in other 
systems with dipolar inter-Skyrmion forces \cite{Jaykka.Speight:10,
Garaud.Sellin.ea:14,Agterberg.Babaev.ea:14} show, for example, that hexagonal 
symmetry is unfavorable for the Skyrmion lattices. Moreover, the long-range 
dipolar interaction can also result in long-range attractive interaction 
between Skyrmions and boundaries of a superconductor, thus suggesting possible 
abundance of topological defects near boundaries in a weak applied magnetic field.

Bound states of nonoverlapping HQVs are coreless defects that can be 
called Skyrmions, and the reason for that terminology is that they exhibit 
additional topological properties, as compared to singular vortices. These 
can be seen by introducing the unit vector $\bf n$ defined as the projection 
of the superconducting degrees of freedom $\eta^\dagger=(\Delta_+^*,\Delta_-^*)$ 
onto spin-$1/2$ Pauli matrices $\tilde{\bs\sigma}$, as 
${\bf n}=\eta^\dagger\tilde{\bs \sigma}\eta/\eta^\dagger\eta$. That is, 
the $x$ and $y$ components of the vector ${\bf n}$ depend on the phase 
difference, while the $z$ component is determined by the ratio of the moduli 
of the complex fields.
The associated projection is a map from the one-point compactification of 
the plane ($\Real^2\cup\{\infty\}\simeq S^2 $) onto the two-sphere target 
space spanned by $\bf n$. That is ${\bf n}: S^2\to S^2$, which is classified 
by the homotopy class $\pi_2(S^2)\in\mathrm{Z}$. This defines the integer-valued 
$\groupCP{1}$ topological invariant, as 
\Equation{Charge}{
   \Q({\bf n})=\frac{1}{4\pi} \int_{\Real^2}
   {\bf n}\cdot\partial_x {\bf n}\times \partial_y {\bf n}\,\,
  dxdy\,.
}
If $\eta\neq0$ everywhere (coreless vortex), $\Q$ is an integer number. 
In a way, $\Q$ counts the number of times the texture of $\bf n$ covers 
the target two-sphere. 

Figure \ref{Fig:Texture} shows the texture of the unit vector $\bf n$ that 
corresponds to the vortices in the nematic phase. The left panel corresponds
to the single-quantum vortex displayed in \Figref{Fig:Nematic}. It illustrates 
that the unit vector ${\bf n}$ wraps the target two-sphere (once), thus 
implying this configuration has unit Skyrmionic charge $\Q=1$. The right panel 
shows the texture corresponding to the bound state of \Figref{Fig:Nematic2} 
that originates in long-range dipolelike forces. 
Note that this illustrates the dipole nature of the long-range interaction. 
Indeed the pair of Skyrmions alternates the north (red) and south (blue) poles 
of the target sphere.

It is worth noting that various coreless vortices were also considered to 
exist in a number of models of multicomponent superconductivity. There are, 
however, substantial differences in the structure and properties of these 
solutions, and they should have distinct experimental manifestations. 
For example, in the framework of various models of $p$-wave superconductors, 
it was advocated that multiquanta coreless vortices may be favored over 
single-quanta singular vortices  \cite{Sauls.Eschrig:09,Ichioka.Machida.ea:12,
knigavko1999magnetic}. The recent numerical studies show that two-quanta 
Skyrmions in the GL models of a chiral $p$-wave superconductor are energetically 
favored, and hence single-quanta Skyrmions do not form in the ground state 
in an external field \cite{Garaud.Babaev:15a,Garaud.Babaev.ea:16,
FernandezBecerra.Sardella.ea:16,Zhang.Becerra.ea:16,Becerra.Milosevic:17}. 
Different types of chiral Skyrmions were also discussed for $s+is$ 
superconducting states \cite{Garaud.Carlstrom.ea:11,Garaud.Carlstrom.ea:13,
Garaud.Babaev:15a}. The structure of chiral Skyrmions is significantly 
dissimilar compared to the nematic Skyrmions since the former have fractional 
vortices and magnetic flux pinned on domain walls between different time-reversal 
symmetry broken states \cite{Garaud.Babaev:14}; thus, they have very different 
magnetic field configurations.

The object that we find here should be interesting from the viewpoint of 
electronic states. Indeed, HQVs are known to possess Majorana modes. The spatial 
separation between two HQVs in the Skyrmion implies that individual HQVs may 
be rather easily stabilized in a mesoscopic sample. 

In conclusion, we discussed the topological excitations in nematic superconductors.
We showed that the topological excitations are nematic Skyrmions, each of which 
can be viewed as a bound state of two spatially separated half-quantum vortices. 
The nematic Skyrmions have orientation-dependent dipolar attractive forces and 
form multiquanta bound states, which could be expected to have clear experimental 
signatures in structure formation. 
Moreover, being coreless, the Skyrmions are expected to have unusual electronic 
core-state properties that could allow the identification of Skyrmionic 
states using scanning tunneling microscopy.

\begin{acknowledgments}
The work was supported by the Swedish Research Council 
Grant No. 642-2013-7837 and by Goran Gustafsson Foundation for Research in Natural Sciences and Medicine. 
The computations were performed on resources provided by 
the Swedish National Infrastructure for Computing (SNIC) 
at National Supercomputer Center at Link\"oping, Sweden.
\end{acknowledgments}

%


\onecolumngrid
\begin{center}
\rule{0.38\linewidth}{1pt}\\
\vspace{-0.37cm}\rule{0.49\linewidth}{1pt}
\end{center}
\setcounter{section}{0}
\setcounter{equation}{0}
\renewcommand{\theequation}{\Alph{subsection}.\arabic{equation}}

\section*{Supplemental Material to \texorpdfstring{\\}{}
"Nematic Skyrmions in odd-parity superconductors"}

In the Supplemental Material, we discuss the spectrum of quasiparticles and derive the Ginzburg-Landau free energy functional of the nematic superconductor introduced in the main text. We also evaluate expression for the spin polarization in the nematic superconductor. 

\subsection{Main definitions}
We introduce the model of the superconductivity in the doped Dirac material. As a particular example we consider theoretical model of a doped $\mathrm{Bi_2Se_3}$.  This is a spatial inversion and time reversal symmetric layered material with two orbitals within single quintuple layer. For concreteness, we assume that $z$-axis is normal to the plane of the layers.  The low energy excitations in the presence of the magnetic field can be described by the Hamiltonian $\mathcal{H}=\int d^3r \Psi^{\dag}(\mathbf{r})H(\mathbf{r})\Psi(\mathbf{r})$, where
\begin{eqnarray}\label{Main}
H(\mathbf{r})=
v\tau_z\left[\bsigma \times \bigg(-i\bnabla-\frac{e}{c}\mathbf{A}(\mathbf{r})\bigg)\right]\cdot \hat{z}
+
v_z\tau_y\bigg(-i\partial_z-\frac{e}{c}\mathbf{A}(\mathbf{r})\bigg)+m\tau_x +(\mathbf{h}+\mathbf{m}\tau_x)\cdot\bsigma,
\end{eqnarray}
in which $v$ and $v_z$ are the in-plane and out of plane components of the Fermi velocity, $|m|$ is the mass of the Dirac fermion describing electron tunneling between the orbitals of the quintuple layer, $e<0$ is the charge of electron, $\bsigma$ and $\btau$ are the Pauli matrices describing the spin and orbital degrees of freedom, such that
\begin{equation}
\Psi(\mathbf{r})=(\Psi_{\uparrow,1}(\mathbf{r}),\Psi_{\downarrow,1}(\mathbf{r}),\Psi_{\uparrow,2}(\mathbf{r}),\Psi_{\downarrow,2}(\mathbf{r}))^{\mathrm{T}}.
\end{equation}
The presence of two orbitals with opposite sign of the coefficient, $v$, supports inversion symmetry in the system.
The Zeeman effect of the external magnetic field results in the spin polarization of electrons in the orbital $\mathbf{h}\cdot \bsigma$ and spin polarized inter-orbital coupling $\mathbf{m}\cdot \bsigma\tau_x$.  Although we have written these contributions in the Hamiltonian for completeness we will ignore Zeeman effect further compared to the orbital effect of the magnetic field. 
The spectrum of particles in the absence of magnetic field is given by
\begin{equation}
E_{\pm}(\mathbf{p})=\pm\{v^2p_{\perp}^2+v_z^2p_z^2+m^2\}^{1/2}.
\end{equation}
This spectrum is anisotropic and has a gap $2m$ at $p\equiv |\mathbf{p}|=0$.

We consider phonon mediated attractive interaction, which can lead to the instabilities in the inter-orbital and intra-orbital couplings. Part of the Hamiltonian, which describes attractive electron-electron interaction is given by 
\begin{equation}
\mathcal{H}_{\mathrm{int}} = -\sum_{\tau}\int d^3r \Psi^{\dag}_{\downarrow\tau}(\mathbf{r})\Psi^{\dag}_{\uparrow\tau}(\mathbf{r})U_{\tau} \Psi_{\uparrow\tau}(\mathbf{r})\Psi_{\downarrow\tau}(\mathbf{r}) -\sum_{\sigma,\sigma'}\int d^3r \Psi^{\dag}_{\sigma 1}(\mathbf{r})\Psi^{\dag}_{\sigma' 2}(\mathbf{r})\lambda \Psi_{\sigma' 2}(\mathbf{r})\Psi_{\sigma 1}(\mathbf{r}) 
\end{equation}
We note that we consider the zero harmonic of the electron-phonon interaction potential in the intra-orbital, $U_{\tau}>0$, and inter-orbital channels, $\lambda>0$, which shall give higher temperature of the superconductor-metal phase transition compared to higher harmonics.
It can be shown that the spectrum of the Bogolubov quasiparticles in the case of the inter-orbital spin-singlet pairing contains nodal line. We ignore this coupling mechanism here, since such a pairing is energetically unfavorable compared to the intra-orbital spin-singlet and inter-orbital spin-triplet pairings, which are either fully gapped or contain point nodes in the spectrum. 

Thus, taking into account only orbital effects of the magnetic field, the BCS Hamiltonian is given by
\begin{eqnarray}
\mathcal{H}_{\mathrm{BCS}}&=&\int d^3r \Psi^{\dag}(\mathbf{r})[v\tau_z\left[\bsigma \times \bigg(-i\bnabla-\frac{e}{c}\mathbf{A}(\mathbf{r})\bigg)\right]\cdot \hat{z}
+
v_z\tau_y\bigg(-i\partial_z-\frac{e}{c}\mathbf{A}(\mathbf{r})\bigg)+m\tau_x -\mu]\Psi(\mathbf{r})\\
&-&\sum_{\tau}\int d^3r\left[\Delta_{\tau}^{+}(\mathbf{r})\Psi_{\uparrow\tau}(\mathbf{r})\Psi_{\downarrow\tau}(\mathbf{r})+ \Psi^{\dag}_{\downarrow\tau}(\mathbf{r})\Psi^{\dag}_{\uparrow\tau}(\mathbf{r})\Delta_{\tau}(\mathbf{r})-\frac{|\Delta_\tau(\mathbf{r})|^2}{U_{\tau}}\right]\\
&-& \sum_{\sigma,\sigma'}
\int d^3r\bigg[ \Delta^{+}_{\sigma\sigma'}(\mathbf{r})\Psi_{\sigma' 2}(\mathbf{r})\Psi_{\sigma 1}(\mathbf{r})+\Psi^{\dag}_{\sigma 1}(\mathbf{r})\Psi^{\dag}_{\sigma' 2}(\mathbf{r})\Delta_{\sigma'\sigma}(\mathbf{r})-\frac{|\Delta_{\sigma\sigma'}(\mathbf{r})|^2}{\lambda}\bigg]
\end{eqnarray} 
Here second line describes intra-orbital spin-singlet pairing: $\Delta_{\tau}(\mathbf{r}) = U_{\tau}\langle \Psi_{\uparrow\tau}(\mathbf{r}) \Psi_{\downarrow \tau}(\mathbf{r})\rangle $, $\Delta_{\tau}^{+}(\mathbf{r}) = U_{\tau}\langle \Psi^{\dag}_{\downarrow \tau}(\mathbf{r})\Psi^{\dag}_{\uparrow\tau}(\mathbf{r}) \rangle $, third line describes inter-orbital pairing: $\Delta_{\sigma'\sigma}(\mathbf{r}) = \lambda\langle \Psi_{\sigma' 2}(\mathbf{r}) \Psi_{\sigma 1}(\mathbf{r})\rangle $, $\Delta_{\sigma\sigma'}^{+}(\mathbf{r}) = \lambda\langle \Psi^{\dag}_{\sigma 1}(\mathbf{r})\Psi^{\dag}_{\sigma' 2}(\mathbf{r}) \rangle $, which might contain both spin-singlet and spin-triplet components; $\mu$ is the Fermi energy. The BCS Hamiltonian can be rewritten as
\begin{equation}
\mathcal{H}_{\textrm{BCS}}=\frac{1}{2}\int d^3r \Phi^{\dag}(\mathbf{r})H_{\textrm{BCS}}(\mathbf{r})\Phi(\mathbf{r})+ \int d^3r \bigg[\sum_{\tau}\frac{|\Delta_\tau(\mathbf{r})|^2}{U_{\tau}} + \sum_{\sigma,\sigma'}\frac{|\Delta_{\sigma\sigma'}(\mathbf{r})|^2}{\lambda}\bigg]
\end{equation}
where we have performed unitary transformation and introduced new operators $\Phi^{\dag}(\mathbf{r})=(\Psi^{\dag}(\mathbf{r}),
\Psi^{\textrm{T}}(\mathbf{r})(-i\sigma_y))$, which are written in the Nambu space. The Hamiltonian is given by
\begin{equation}
H_{\textrm{BCS}}(\mathbf{r}) = \bigg[
 \begin{matrix}
  H(\mathbf{r}) & \Delta(\mathbf{r}) \\
  \Delta^{\dag}(\mathbf{r}) & -\sigma_y H^{*}(\mathbf{r})\sigma_y
 \end{matrix}
\bigg],
\end{equation}
We note that here each block is a $4\times 4$ matrix and that Hamiltonian satisfies an equality
\begin{equation}
\sigma_yH^*(\mathbf{r})\sigma_y = H(\mathbf{r})|_{\mathbf{A}(\mathbf{r})\rightarrow -\mathbf{A}(\mathbf{r})},
\end{equation}
Order parameters can be generally written as
\begin{equation}
\Delta(\mathbf{r}) =   \frac{\Delta_1(\mathbf{r})+\Delta_2(\mathbf{r})}{2}+ \frac{\Delta_1(\mathbf{r})-\Delta_2(\mathbf{r})}{2}\tau_z  +  \bigg[
 \begin{matrix}
  \Delta_{\uparrow\downarrow}(\mathbf{r}) & \Delta_{\downarrow\downarrow}(\mathbf{r}) \\
  -\Delta_{\uparrow\uparrow}(\mathbf{r}) & -\Delta_{\downarrow\uparrow}(\mathbf{r})
 \end{matrix}
\bigg]\frac{\tau_x+i\tau_y}{2} -  \bigg[
 \begin{matrix}
  \Delta_{\downarrow\uparrow}(\mathbf{r}) & \Delta_{\downarrow\downarrow}(\mathbf{r}) \\
  -\Delta_{\uparrow\uparrow}(\mathbf{r}) & -\Delta_{\uparrow\downarrow}(\mathbf{r})
 \end{matrix}
\bigg]\frac{\tau_x-i\tau_y}{2}.
\end{equation}
Neglecting spin-singlet inter-orbital coupling one can parametrize superconducting order parameters as
\begin{equation}
\Delta(\mathbf{r}) =   \frac{\Delta_1(\mathbf{r})+\Delta_2(\mathbf{r})}{2}+ \frac{\Delta_1(\mathbf{r})-\Delta_2(\mathbf{r})}{2}\tau_z  +  \bsigma\cdot\boldsymbol{\Delta}(\mathbf{r}) \tau_y,
\end{equation}
where we introduce a vector $\boldsymbol{\Delta}(\mathbf{r}) = (\Delta_x(\mathbf{r}),\Delta_y(\mathbf{r}),\Delta_z(\mathbf{r}))$, such that
\begin{equation}
\Delta_x(\mathbf{r}) = -\frac{i}{2}(
\Delta_{\uparrow\uparrow}(\mathbf{r})-\Delta_{\downarrow\downarrow}(\mathbf{r})),~~\Delta_y(\mathbf{r}) = -\frac{1}{2}(\Delta_{\uparrow\uparrow}(\mathbf{r})+\Delta_{\downarrow\downarrow}(\mathbf{r}))~~\Delta_z(\mathbf{r}) = \frac{1}{2}(\Delta_{\uparrow\downarrow}(\mathbf{r})+\Delta_{\downarrow\uparrow}(\mathbf{r})) 
\end{equation}
and assume that the inter-orbital spin-singlet pairing $\propto \Delta_{\uparrow\downarrow}(\mathbf{r})-\Delta_{\downarrow\uparrow}(\mathbf{r})=0$ vanishes.

\subsection{Eigenvalues of the BCS Hamiltonian}
Here we consider the spectrum of quasiparticles in the Dirac superconductor in few limiting cases. We set external magnetic field to zero and consider coordinate independent order parameters. The BCS Hamiltonian in the momentum representation is given by
\begin{equation}
H_{\textrm{BCS}}(\mathbf{k}) = \bigg[
 \begin{matrix}
  H(\mathbf{k}) & \Delta\\
  \Delta^{\dag} & -H(\mathbf{k})
 \end{matrix}
\bigg],
\end{equation}
where 
\begin{eqnarray}
 H(\mathbf{k}) &=& v\tau_z\left[\bsigma \times \mathbf{k}\right]\cdot \hat{z}
+ v_z\tau_yk_z+m\tau_x -\mu,\\
\Delta &=&   \frac{\Delta_1+\Delta_2}{2}+ \frac{\Delta_1-\Delta_2}{2}\tau_z  +  \bsigma\cdot\boldsymbol{\Delta} \tau_y.
\end{eqnarray}
We will assume that the Fermi level is in the conduction band $\mu>|m|$ and for concreteness consider several representative cases of the order parameter. 
\\
1. Consider the intra-orbital pairing and assume $\Delta_1=\Delta e^{i\phi}, \Delta_2=\Delta$, where $\phi$ is the phase difference of the order parameters in two orbitals.
The spectrum of quasiparticles is described by
\begin{eqnarray}
\mathcal{E}^2(\mathbf{p}) = (E_{+}(\mathbf{p}) -\mu)^2 + |\Delta|^2 \left[ 1-\frac{(1-\cos{\phi})(m^2+v_z^2p_z^2)}{2\mu^2}\right],
\end{eqnarray}
which is gapped if the phase difference is $\phi=0$, while it has two point nodes at $\mathbf{p}_{\pm}=(0,0, \pm \sqrt{\mu^2-m^2}/v_z)$ if the phase difference $\phi=\pi$.\\
2. Consider inter-orbital spin-triplet pairing in few limiting cases. In the case $\mathbf{\Delta}=(0,0,\Delta_z)$ the spectrum is gapped
\begin{eqnarray}
\mathcal{E}^2(\mathbf{p}) = (E_{+}(\mathbf{p}) - \mu)^2 
+|\Delta_z|^2(1-m^2/\mu^2).
\end{eqnarray} \\
3. In the case when $\boldsymbol{\Delta}=(\Delta_{x},\Delta_{y},0)$ and when both components $\Delta_{x,y}$ are real,the low energy excitations at the Fermi level are described by
\begin{eqnarray}
\mathcal{E}^2(\mathbf{p}) = (E_{+}(\mathbf{p}) - \mu)^2 
+(\Delta_x^2+\Delta_y^2) \bigg(1-\frac{m^2}{\mu^2}\bigg) -\frac{(vp_x\Delta_y-vp_y\Delta_x)^2}{\mu^2}
\end{eqnarray}
There exist point nodes in the plane $p_z=0$. For example, if $\boldsymbol{\Delta}= \Delta(\cos(\theta),\sin(\theta),0)$, then the positions of the nodes are determined by the solution of equation 
\begin{equation}
\sin^2(\phi-\theta)=\frac{\mu^2-m^2}{v^2p_{\perp,F}^2},
\end{equation}
 where we set $p_{x}=p_{\perp,F}\cos(\phi), p_{y}=p_{\perp,F}\sin(\phi)$ for the components of the in-plane momentum at the Fermi level.

\subsection{Spin polarization: magnetoelectric effect in nematic superconductor}
In this section we will derive expression for the quasiparticle spin polarization in the superconductor in the absence of the magnetic field. The electron spin density $\mathbf{S}(\mathbf{r})$ can be written as follows:
\begin{equation}
\mathbf{S}(\mathbf{r}) = \lim_{\mathbf{r}'\rightarrow \mathbf{r}}\textrm{Tr}~T\sum_n \frac{\bsigma}{2}G(\mathbf{r},\mathbf{r}',\omega_n),
\end{equation}
where $G(\mathbf{r},\mathbf{r}',\omega_n)$ is the Green function in Matsubara representation, $T$ is the temperature, and trace is taken over spin and layer degrees of freedom. One can also define pseudospin
\begin{equation}
\mathbf{P}_i(\mathbf{r}) = \lim_{\mathbf{r}'\rightarrow \mathbf{r}}\textrm{Tr}~T\sum_n \frac{\bsigma}{2}\tau_{i}G(\mathbf{r},\mathbf{r}',\omega_n),
\end{equation}
which describes relative spin polarization in different orbitals. One has two coupled equations for normal and anomalous Green functions
\begin{eqnarray}\nonumber
&[&i\omega_n-H(\mathbf{r})] G(\mathbf{r},\mathbf{r}',\omega_n)  = \delta(\mathbf{r}-\mathbf{r}') - \Delta(\mathbf{r}) \bar{F}(\mathbf{r},\mathbf{r}',\omega_n),\\
&[&i\omega_n +\sigma_y H^{*}(\mathbf{r})\sigma_y]\bar{F}(\mathbf{r},\mathbf{r}',\omega_n) = -\Delta^{\dag}(\mathbf{r})G(\mathbf{r},\mathbf{r}',\omega_n),
\end{eqnarray}
supplemented by the self-consistency equation for the order parameter
\begin{equation}
\Delta(\mathbf{r})=\lambda T\sum_n\lim_{\mathbf{r}'\rightarrow \mathbf{r}}F(\mathbf{r},\mathbf{r}',\omega_n),
\end{equation}
where $\lambda$ is the interaction constant.
In order to find expression for the Green function $G(\mathbf{r},\mathbf{r}',\omega_n) $, we expand in powers of $\Delta(\mathbf{r})$ up to the second order. Thus, solving equations for the Green functions perturbatively in the order parameter
\begin{eqnarray}
G(\mathbf{r},\mathbf{r}',\omega_n) &=& G_0(\mathbf{r},\mathbf{r}',\omega_n) 
- \int d\mathbf{r}_1 G_0(\mathbf{r},\mathbf{r}_1,\omega_n)\Delta(\mathbf{r}_1)\bar{F}(\mathbf{r}_1,\mathbf{r}',\omega_n),
\\\nonumber
\bar{F}(\mathbf{r},\mathbf{r}',\omega_n) &=& - \int d\mathbf{r}_1 \bar{G}_0(\mathbf{r},\mathbf{r}_1,\omega_n)\Delta^{\dag}(\mathbf{r}_1)G(\mathbf{r}_1,\mathbf{r}',\omega_n)
\end{eqnarray}
we obtain
\begin{eqnarray}
G(\mathbf{r},\mathbf{r}',\omega_n) &=& G_0(\mathbf{r},\mathbf{r}',\omega_n)+\int d^3r_1d^3r_2G_0(\mathbf{r},\mathbf{r}_1,\omega_n) \Delta(\mathbf{r}_1)\bar{G}_0(\mathbf{r}_1,\mathbf{r}_2,\omega_n)\Delta^{\dag}(\mathbf{r}_2)
G_0(\mathbf{r}_2,\mathbf{r}',\omega_n),\\
\end{eqnarray}
where Green functions $G_0(\mathbf{r},\mathbf{r}',\omega_n)$ and $\bar{G}_0(\mathbf{r},\mathbf{r}',\omega_n)$ correspond to the case of the absence of the superconductivity and satisfy equations
\begin{eqnarray}\nonumber
&[&i\omega_n-H(\mathbf{r})]G_0(\mathbf{r},\mathbf{r}',\omega_n)=\delta(\mathbf{r}-\mathbf{r}'),\\
&[&i\omega_n+\sigma_y H^{*}(\mathbf{r})\sigma_y]\bar{G}_0(\mathbf{r},\mathbf{r}',\omega_n)=\delta(\mathbf{r}-\mathbf{r}').
\end{eqnarray}
We can rewrite expression for the spin polarization in the form
\begin{eqnarray}
\mathbf{S}(\mathbf{r}) &=& \lim_{\mathbf{r}'\rightarrow \mathbf{r}}\textrm{Tr}~T\sum_n \frac{\bsigma}{2}
\int d^3r_1d^3r_2G_0(\mathbf{r},\mathbf{r}_1,\omega_n) \Delta(\mathbf{r}_1)\bar{G}_0(\mathbf{r}_1,\mathbf{r}_2,\omega_n)\Delta^{\dag}(\mathbf{r}_2)
G_0(\mathbf{r}_2,\mathbf{r}',\omega_n),
\end{eqnarray}

We are interested in the terms linear to the gradients of the order parameter.
\begin{eqnarray}\nonumber
\mathbf{S}(\mathbf{r}) &=& \textrm{Tr}~T\sum_n \frac{\bsigma}{2}
\int \frac{d^3p}{(2\pi)^3}
\bigg[G_0(\mathbf{p},\omega_n)
\Delta(\mathbf{r})\bar{G}_0(\mathbf{p},\omega_n) \Delta^{\dag}(\mathbf{r})
G_0(\mathbf{p},\omega_n)
+\frac{i}{2}G_0(\mathbf{p},\omega_n)
\bigg\{ \left[\bnabla\Delta(\mathbf{r})\right]\cdot\frac{\partial}{\partial \mathbf{p}}G_0(\mathbf{p},-\omega_n)\Delta^{\dag}(\mathbf{r})\\
&-&  \Delta(\mathbf{r})\bigg[\frac{\partial}{\partial \mathbf{p}}G_0(\mathbf{p},-\omega_n)\bigg]\cdot\bnabla\Delta^{\dag}(\mathbf{r})
\bigg\}G_0(\mathbf{p},\omega_n)\bigg],
\end{eqnarray}
where we are using momentum representation of the Green function
\begin{equation}
G_0(\mathbf{p},\omega_n) = \int d^3r e^{-i\mathbf{p}\cdot(\mathbf{r}-\mathbf{r}')}G_0(\mathbf{r},\mathbf{r}',\omega_n),
\end{equation}
which is given by
\begin{equation}
G_0(\mathbf{p},\omega_n) = \frac{1}{2}\sum_{s=\pm}\frac{D_s(\mathbf{p})}{i\omega_n+\mu-E_{s}(\mathbf{p})},
\end{equation}
where $E_{\pm}(\mathbf{p}) = \pm\{v^2p_{\perp}^2+v_z^2p_z^2+m^2\}^{1/2}$ is the spectrum of particles in the non-superconducting state and
\begin{equation}
D_s(\mathbf{p})=1+\{v[\bsigma\times\mathbf{p}]\cdot \hat{z}\tau_z+v_zk_z\tau_y+m\tau_x\}E^{-1}_s(\mathbf{p})
\end{equation}
is the projector operator on electron or hole bands, which are defined by $s=\pm$. Operator $D_s(\mathbf{p})$ satisfies an equality 
$
D_s^2(\mathbf{p})/4=D_s(\mathbf{p})/2.
$
We also observe that 
\begin{equation}
\bar{G}_0(\mathbf{p},\omega_n)=-G_0(\mathbf{p},-\omega_n)
\end{equation}
and
\begin{eqnarray}\label{expand_velocity}
\frac{\partial}{\partial \mathbf{p}}G_0(\mathbf{p},-\omega_n)\cdot\bnabla =
G_0(\mathbf{p},-\omega_n) \{v_z\tau_y\nabla_z+v[\bsigma\times \bnabla]\cdot\hat{z}\tau_z\}G_0(\mathbf{p},-\omega_n) .
\end{eqnarray}

Now we consider the case when the Fermi energy is in the conduction band, $\mu>|m|$, in which it is enough to take into account contribution from the band with $s=+$. Noting that
\begin{equation}
\Delta(\mathbf{r}) =   \frac{\Delta_1(\mathbf{r})+\Delta_2(\mathbf{r})}{2}+ \frac{\Delta_1(\mathbf{r})-\Delta_2(\mathbf{r})}{2}\tau_z  +  \bsigma\cdot\boldsymbol{\Delta}(\mathbf{r}) \tau_y,
\end{equation}
we find expression for the spin polarization
\begin{eqnarray}
\mathbf{S}(\mathbf{r}) &=& \frac{T}{2}\sum_n\int\frac{d^3p}{(2\pi)^3}\frac{v^2p_{\perp}^2+2v_z^2p_z^2}{E_{+}^2(\mathbf{p})} \frac{i[\boldsymbol{\Delta}(\mathbf{r})\times\boldsymbol{\Delta}^{*}(\mathbf{r})]}{(i\omega_n+\mu-E_{+}(\mathbf{p}))^2(i\omega_n-\mu+E_{+}(\mathbf{p}))}
\\\nonumber
&+&\frac{vT}{16}\sum_{n}\int\frac{d^3p}{(2\pi)^3}\frac{v^2p_{\perp}^2}{E_{+}^2(\mathbf{p})} \frac{\textrm{Im}(\Delta_1^{*}(\mathbf{r})[\hat{z}\times\bnabla \Delta_1(\mathbf{r})]-\Delta_2^{*}(\mathbf{r})[\hat{z}\times\bnabla \Delta_2(\mathbf{r})])}{[\omega_n^2+(E_{+}(\mathbf{p})-\mu)^2]^2}.
\end{eqnarray}
 We observe that only second term on the right hand side of Eq.~\ref{expand_velocity} contributes to the spin polarization through the first gradients of the order parameter. Consider simple current carrying expression for the order parameters $\Delta_{1,2}(\mathbf{r}) = |\Delta|e^{i\phi_{1,2}(\mathbf{r})}$. We finally obtain expression for the spin polarization in the Dirac superconductor
\begin{eqnarray}
\mathbf{S}(\mathbf{r}) &=& \frac{T}{2}\sum_n\int\frac{d^3p}{(2\pi)^3}\frac{v^2p_{\perp}^2+2v_z^2p_z^2}{E_{+}^2(\mathbf{p})} \frac{i[\boldsymbol{\Delta}(\mathbf{r})\times\boldsymbol{\Delta}^{*}(\mathbf{r})]}{(i\omega_n+\mu-E_{+}(\mathbf{p}))^2(i\omega_n-\mu+E_{+}(\mathbf{p}))}
\\\nonumber
&+& \frac{v|\Delta|^2}{16}T\sum_{n}\int\frac{d^3p}{(2\pi)^3}\frac{v^2p_{\perp}^2}{E_{+}^2(\mathbf{p})}
 \frac{[\hat{z}\times\bnabla (\phi_{1}(\mathbf{r})-\phi_{2}(\mathbf{r}))]}{[\omega_n^2+(E_{+}(\mathbf{p})-\mu)^2]^2} 
\end{eqnarray}

To summarize this section, \\
1. Finite term $\propto|\boldsymbol{\Delta}(\mathbf{r})\times\boldsymbol{\Delta}^{*}(\mathbf{r})|\neq 0$ describes spin polarization in the spontaneously time reversal symmetry broken state of the superconductor. Nontrivial pseudospin polarization $\mathbf{P}_z(\mathbf{r})$ shows up in the first gradients of the order parameter in both chiral and nematic cases. Interestingly, pseudospin polarization results in the antiferromagnetic spin orientation in two orbitals even in the nematic case.\\
2. Finite gradient of the phase difference of the order parameters in two orbitals $\bnabla[\phi_{1}(\mathbf{r})-\phi_{2}(\mathbf{r})]\neq 0$ breaks time reversal symmetry and thus leads to the spin polarization in the superconductor, $\propto [\hat{z}\times\bnabla \{\phi_{1}(\mathbf{r})-\phi_{2}(\mathbf{r})\}]$. This is the well known magneto-electric effect in
superconductors. Contrary, if $\bnabla[\phi_{1}(\mathbf{r})-\phi_{2}(\mathbf{r})]= 0$, while $\bnabla \phi_{1,2}(\mathbf{r})\neq 0$, then there exists finite pseudo-spin polarization, which corresponds to the antiferromagnetic in-plane spin orientation. Since there might be in-plane easy axis, anisotropy of the pseudospin polarization is expected.

\subsection{Derivation of Ginzburg-Landau equation}
In this section we will derive Ginzburg-Landau equation and Ginzburg-Landau functional density for the case of inter-orbital spin-triplet pairing
\begin{equation}
\Delta(\mathbf{r}) = \bsigma\cdot\boldsymbol{\Delta}(\mathbf{r}) \tau_y.
\end{equation}
Due to strong anisotropy of the spectrum, it is convenient to separate the case $\boldsymbol{\Delta}(\mathbf{r}) = (\Delta_x(\mathbf{r}),\Delta_y(\mathbf{r}),0)$ from  $\boldsymbol{\Delta}(\mathbf{r}) = (0,0,\Delta_z(\mathbf{r}))$. In the latter case the Ginzburg-Landau free energy functional has standard expression for the single order parameter. Thus, in what follows we set $\Delta_z =0$ and
consider a vector $\boldsymbol{\Delta}(\mathbf{r})=(\Delta_x(\mathbf{r}),\Delta_y(\mathbf{r}),0)$, which allows to write 
\begin{equation}
\Delta(\mathbf{r}) = [\sigma_x\Delta_x(\mathbf{r})+\sigma_y\Delta_y(\mathbf{r})] \tau_y.
\end{equation}

We start from the self-consistency equation for the order parameter
\begin{equation}
\Delta^{\dag} (\mathbf{r}) = \lambda T\sum_n \lim_{\mathbf{r}'\rightarrow \mathbf{r}} \bar{F}(\mathbf{r},\mathbf{r}',\omega_n),
\end{equation}
where $\lambda$ is the interaction constant. Anomalous Green function is defined by
\begin{eqnarray}
\bar{F}(\mathbf{r},\mathbf{r}',\omega_n) &=& - \int d\mathbf{r}_1 \bar{G}_0(\mathbf{r},\mathbf{r}_1,\omega_n)\Delta^{\dag}(\mathbf{r}_1)G_0(\mathbf{r}_1,\mathbf{r}',\omega_n)
\\ \nonumber
&-& \int d\mathbf{r}_1 d\mathbf{r}_2 d\mathbf{r}_3
\bar{G}_0(\mathbf{r},\mathbf{r}_1,\omega_n)\Delta^{\dag}(\mathbf{r}_1)G_0(\mathbf{r}_1,\mathbf{r}_2,\omega_n)
\Delta(\mathbf{r}_2)
\bar{G}_0(\mathbf{r}_2,\mathbf{r}_3,\omega_n)
\Delta^{\dag}(\mathbf{r}_3)G_0(\mathbf{r}_3,\mathbf{r}',\omega_n)+...
\end{eqnarray}

Thus in momentum representation (we will be include orbital effect of the magnetic field in the final expressions by introducing the gauge invariant derivatives) we obtain
\begin{eqnarray}
\Delta^{\dag} (\mathbf{q}) &=& -\lambda T\sum_n \int (dp)\bar{G}_0(\mathbf{p},\omega_n)\Delta^{\dag} (\mathbf{q})G_0(\mathbf{p}-\mathbf{q},\omega_n)
\\
\nonumber
&-&\lambda T\sum_n \int (dp)\bar{G}_0(\mathbf{p},\omega_n)\Delta^{\dag} (\mathbf{q})G_0(\mathbf{p},\omega_n)\Delta(\mathbf{q})\bar{G}_0(\mathbf{p},\omega_n)\Delta^{\dag} (\mathbf{q})G_0(\mathbf{p},\omega_n)
+...,
\end{eqnarray}
where we have neglected derivatives of the order parameter in the non-linear in powers of the order parameter terms. Using that the Green functions satisfy equality
\begin{equation}
\bar{G}_0(\mathbf{p},\omega_n)=-G_0(\mathbf{p},-\omega_n),
\end{equation}
we obtain
\begin{eqnarray}
\Delta^{\dag} (\mathbf{q}) &=& \lambda T\sum_n \int (dp)G_0(\mathbf{p},-\omega_n)\Delta^{\dag} (\mathbf{q})G_0(\mathbf{p}-\mathbf{q},\omega_n)
\\\nonumber
&-&\lambda T\sum_n \int (dp)G_0(\mathbf{p},-\omega_n)\Delta^{\dag} (\mathbf{q})G_0(\mathbf{p},\omega_n)\Delta(\mathbf{q})G_0(\mathbf{p},-\omega_n)\Delta^{\dag} (\mathbf{q})G_0(\mathbf{p},\omega_n)\\
\nonumber
&-&
\lambda T\sum_n \int (dp)G_0(\mathbf{p},-\omega_n)\Delta^{\dag} (\mathbf{q})G_0(\mathbf{p},\omega_n)\Delta(\mathbf{q})G_0(\mathbf{p},-\omega_n)\Delta^{\dag} (\mathbf{q})G_0(\mathbf{p},\omega_n)
\Delta(\mathbf{q})G_0(\mathbf{p},-\omega_n)\Delta^{\dag} (\mathbf{q})G_0(\mathbf{p},\omega_n).
\end{eqnarray}
Here we also include fifth order term. Since we consider that the Fermi level is in the conduction band and the Fermi energy $\mu>|m|$ is the largest energy scale in the system, we can approximate the Green function as
\begin{equation}
G_0(\mathbf{p},\omega_n) = \frac{1}{2}\frac{D_{+}(\mathbf{p})}{i\omega_n+\mu-E_{+}(\mathbf{p})},
\end{equation}
In the intermediate derivations one obtains terms which contain gradients of the Green function over momentum $\propto\partial_{\mathbf{p}}G_0(\mathbf{p},\omega_n)$ under the integral over momentum. The natural limit of small temperature of the superconductor-metal phase transition $T_c/\mu\ll 1$, allows to approximate these terms as $\frac{D_{+}(\mathbf{p})}{2}\partial_{\mathbf{p}} \frac{1}{i\omega_n+\mu-E_{+}(\mathbf{p})}$. 

We derive Ginzburg-Landau equation for the spin-triplet inter-orbital order parameter $\boldsymbol{\Delta}(\mathbf{r}) = (\Delta_x(\mathbf{r}),\Delta_y(\mathbf{r}),0)$ in the form
\begin{eqnarray}\nonumber
\bigg[\alpha&+&\beta_1 (D_x^2+D_y^2) + \beta_3D_z^2 \pm\beta_2(D_x^2-D_y^2)\bigg]\Delta_{x,y}+\beta_2(D_xD_y+D_yD_x)\Delta_{y,x}\\
&+&\gamma_1\Delta_{x,y}\left(|\Delta_x|^2+|\Delta_y|^2\right) -\gamma_2\Delta_{x,y}^*\left(\Delta_x^2+\Delta_y^2\right)+O(\Delta^5)=0,
\end{eqnarray}
where
\begin{equation}
D_n = -i\partial_n -\frac{2e}{c}A_n;~~n=x,y,z; ~e<0.
\end{equation} 
Coefficients are given by:
\begin{eqnarray}
\alpha &=& 1-\frac{\lambda T}{4}\sum_n\int\frac{(dp)}{E^{2}_{+}(\mathbf{p})(\omega_n^2+\xi^2)}[2v_z^2p_z^2+v^2p_{\perp}^2]
\\
\beta_1 &=& \frac{\lambda T}{16}\sum_n\int\frac{(dp)}{E^{4}_{+}(\mathbf{p})(\omega_n^2+\xi^2)^2}v^4p_{\perp}^2[2v_z^2p_z^2+v^2p_{\perp}^2]
\\
\beta_3 &=& \frac{\lambda T}{8}\sum_n\int\frac{(dp)}{E^{4}_{+}(\mathbf{p})(\omega_n^2+\xi^2)^2}v^4_zp_{z}^2[2v_z^2p_z^2+v^2p_{\perp}^2]
\\
\beta_2 &=& \frac{\lambda T}{16}\sum_n\int\frac{(dp)}{E^{4}_{+}(\mathbf{p})(\omega_n^2+\xi^2)^2}\frac{v^6p_{\perp}^4}{2}
\\
\gamma_1 &=& \frac{\lambda T}{8}\sum_n\int\frac{(dp)}{E^{4}_{+}(\mathbf{p})(\omega_n^2+\xi^2)^2}[8v_z^4p_z^4+v^4p_{\perp}^4+8v_z^2p_z^2v^2p_{\perp}^2]
\\
\gamma_2 &=& \frac{\lambda T}{16}\sum_n\int\frac{(dp)}{E^{4}_{+}(\mathbf{p})(\omega_n^2+\xi^2)^2}[8v_z^4p_z^4-v^4p_{\perp}^4+8v_z^2p_z^2v^2p_{\perp}^2].
\end{eqnarray}
Here $
\xi = E_{+}(\mathbf{p})-\mu,~~ E_{+}(\mathbf{p}) =\sqrt{m^2+v_z^2p_z^2+v^2p_{\perp}^2}$, and $(dp)=\frac{d^3p}{(2\pi)^3}=\frac{p_{\perp}dp_{\perp}}{2\pi}\frac{dp_z}{2\pi}$. The Ginzburg-Landau free energy functional is given by
\begin{eqnarray}
\mathcal{F} &=&\int d^3r\bigg\{ \alpha \boldsymbol{\Delta}\cdot \boldsymbol{\Delta}^{*}  +\beta_{1}\bigg[(D_x\boldsymbol{\Delta})^{*}\cdot (D_x\boldsymbol{\Delta})+(D_y\boldsymbol{\Delta})^{*}\cdot (D_y\boldsymbol{\Delta})\bigg] +\beta_{3}(D_z\boldsymbol{\Delta})^{*}\cdot D_z \boldsymbol{\Delta}
\\\nonumber
&+&\beta_2
\bigg[(D_x\Delta_x)^* (D_x\Delta_x)-(D_y\Delta_x)^*(D_y\Delta_x)+
(D_y\Delta_y)^*(D_y\Delta_y)-(D_x\Delta_y)^*(D_x\Delta_y)\\\nonumber
&+&(D_x\Delta_x)^*(D_y\Delta_y)+(D_y\Delta_x)^*(D_x\Delta_y)
+(D_x\Delta_y)^*(D_y\Delta_x)+(D_y\Delta_y)^*(D_x\Delta_x)\bigg]\\
\nonumber
&+&\frac{\gamma_1-\gamma_2}{2}(\boldsymbol{\Delta}\cdot \boldsymbol{\Delta}^{*})^2
+ \frac{\gamma_2}{2}|\boldsymbol{\Delta}\times \boldsymbol{\Delta}^*|^2+\boldsymbol{\Delta}\cdot \boldsymbol{\Delta}^{*}\bigg[\delta_1(\boldsymbol{\Delta}\cdot \boldsymbol{\Delta}^{*})^2+\delta_2|\boldsymbol{\Delta}\times \boldsymbol{\Delta}^*|^2 \bigg]
\bigg\}
\end{eqnarray}
where $ \boldsymbol{\Delta}\cdot \boldsymbol{\Delta}^{*} =|\Delta_x|^2+|\Delta_y|^2 $ and $\boldsymbol{\Delta}\times \boldsymbol{\Delta}^* = \Delta_x\Delta_y^*-\Delta_x^*\Delta_y$. We have also introduced six order terms with coefficients
\begin{eqnarray}
\delta_1 &=& \frac{\lambda T}{96}\sum_n\int\frac{(dp)(2v_z^2p_z^2+v^2p_{\perp}^2)}{E^{6}_{+}(\mathbf{p})(\omega_n^2+\xi^2)^3}[8v_z^4p_z^4+8v_z^2p_z^2v^2p^2_{\perp}+5v^4p^4_{\perp}]
\\
\delta_2 &=& \frac{\lambda T}{32}\sum_n\int\frac{(dp)(2v_z^2p_z^2+v^2p_{\perp}^2)}{E^{6}_{+}(\mathbf{p})(\omega_n^2+\xi^2)^3}[8v_z^4p_z^4+8v_z^2p_z^2v^2p^2_{\perp}-v^4p^4_{\perp}]
\end{eqnarray}
Few remarks are in order.\\
1. In the superconducting state $\alpha<0$, i.e. at temperatures lower that the critical temperature, $T<T_c$.\\ 
2. Coefficient at the fourth order term
$\gamma_1-\gamma_2>0$, while $\gamma_2$ can be either positive or negative. When $\gamma_2$ is positive the ground state of the superconductor is nematic, when $\gamma_2$ is negative the state is chiral. Note that $\gamma_2$ changes sign from positive at $v_z\sim v$ to negative at $v_z=0$. The latter case describes doped thin film of the three-dimensional topological insulator.\\
3. One observes following relation between coefficients $0<\beta_2/\beta_1\leq 1/2$.\\  
4. We do not include in-plane crystal anisotropy into the model, six order terms do not lift the degeneracy of the nematic state and since they are small compared to the fourth order terms we will ignore them. 
\\
5. To have more insight, consider two limiting cases. First, spherically symmetric case. We set $v=v_z$ and obtain
\begin{eqnarray}
\alpha &=& 1-\frac{\lambda T}{3}\sum_n\int\frac{(dp)}{(\omega_n^2+\xi^2)}\frac{v^2p^2}{m^2+v^2p^2}
\\
\beta_1 &=& v^2\frac{\lambda T}{20}\sum_n\int\frac{(dp)}{(\omega_n^2+\xi^2)^2}\frac{v^4p^4}{(m^2+v^2p^2)^2}\equiv v^2\beta,~~
\beta_3 = \frac{4}{3}v^2\beta,~~
\beta_2 =  \frac{1}{3}v^2\beta
\\
\gamma_1 &=&  8\beta,~~
\gamma_2 =  \frac{8}{3}\beta,~~
\delta_1 = \frac{8\lambda T}{35}\sum_n\int\frac{(dp)}{(\omega_n^2+\xi^2)^3}\frac{v^6p^6}{(m^2+v^2p^2)^3},~~
\delta_2 =  \frac{\delta_1}{2}
\end{eqnarray}
Here $
\xi = E_{+}(\mathbf{p})-\mu,~~ E_{+}(\mathbf{p}) =\sqrt{m^2+v^2p^2},$ and $(dp)=\frac{p^2dp}{2\pi^2}$.
Second, consider cylindrical symmetric case in which we set $v_z=0$.  We find that
\begin{eqnarray}
\alpha &=& 1-\frac{\lambda T}{4}\sum_n\int\frac{(dp)}{(\omega_n^2+\xi^2)}\frac{v^2p^2_{\perp}}{m^2+v^2p_{\perp}^2}
\\
\beta_1 &=& \frac{\lambda T}{16}\sum_n\int\frac{(dp)}{(\omega_n^2+\xi^2)^2}\frac{v^6p^4_{\perp}}{(m^2+v^2p_{\perp}^2)^2}\equiv v^2\tilde{\beta},~~
\beta_3 = 0,~~
\beta_2 =  \frac{v^2\tilde{\beta}}{2}
\\
\gamma_1 &=& \frac{\tilde{\beta}}{2},~~
\gamma_2 =  -\frac{\tilde{\beta}}{4},~~
\delta_1 = \frac{5\lambda T}{96}\sum_n\int\frac{(dp)}{(m^2+v^2p_{\perp}^2)^3}\frac{v^6p_{\perp}^6}{(\omega_n^2+\xi^2)^3},~~
\delta_2 = -\frac{3}{5}\delta_1.
\end{eqnarray}
Here $
\xi = E_{+}(\mathbf{p})-\mu,~~ E_{+}(\mathbf{p}) =\sqrt{m^2+v^2p^2_{\perp}}$, and $(dp)=\frac{p_{\perp}dp_{\perp}}{2\pi}\frac{dp_z}{2\pi}$.
Notice that $\gamma_2$ and $\delta_2$ change sign, which signals for the phase transition between nematic and chiral states of the superconductor.

\subsection{Symmetric form of the Ginzburg-Landau free energy functional}
It is worth to rewrite Ginzburg-Landau free energy functional in the symmetric form. Let us introduce new variables
\begin{equation}
\Delta_{\pm} = (\Delta_x \pm i \Delta_y)/\sqrt{2}, ~~~ D_{\pm} = D_x \pm i D_y
\end{equation} 
Thus, one has
\begin{equation}
\Delta_{x} = (\Delta_+ + \Delta_-)/\sqrt{2},~~ \Delta_{y} = (\Delta_+ - \Delta_-)/\sqrt{2}i.
\end{equation} 
Useful identities
\begin{eqnarray}
\boldsymbol{\Delta}\cdot \boldsymbol{\Delta}^* &=& \Delta_x\Delta_x^*+\Delta_y\Delta_y^* = |\Delta_+|^2+|\Delta_-|^2
\\
|\boldsymbol{\Delta}\times \boldsymbol{\Delta}^*|^2 &=& (\Delta_x\Delta_y^*-\Delta_y\Delta_x^*)(\Delta_y\Delta_x^*-\Delta_x\Delta_y^*)=-(\Delta_x\Delta_y^*-\Delta_y\Delta_x^*)^2=(|\Delta_+|^2-|\Delta_-|^2)^2
\end{eqnarray}
Thus, we rewrite Ginzburg-Landau free energy functional in the symmetric form
\begin{eqnarray}\nonumber
\mathcal{F} &=&\sum_{s=\pm} \int d^3r\bigg\{ \alpha |\Delta_{s}|^2 + \beta_{1}[(D_x\Delta_{s})^*D_x\Delta_{s}+(D_y\Delta_{s})^*D_y\Delta_{s}]+\beta_3 (D_z\Delta_{s})^*D_z\Delta_{s}
\\
&+&
\beta_2(D_{-s}\Delta_{s})^*D_{s}\Delta_{-s}+\frac{\gamma_1}{2}|\Delta_{s}|^4+
\frac{\gamma_1- 2\gamma_2}{2}|\Delta_{s}|^2|\Delta_{-s}|^2\bigg\}
\end{eqnarray}
\subsection{Scaled form of the Ginzburg-Landau free energy functional}
Finally we rewrite Ginzburg-Landau free energy functional in the scaled form for temperatures lower than the temperature of the superconductor-metal phase transition $T<T_c$. We perform standard substitutions:
\begin{eqnarray}
\Delta_s = \psi_s \sqrt{\frac{|\alpha|}{\gamma_1}} .
\end{eqnarray}
We define lengths, which for a single order parameter would correspond to the coherence and magnetic lengths
\begin{eqnarray}
l_{\xi} = \sqrt{\frac{\beta_1}{|\alpha|}},~~ l_m = \frac{c}{2|e|}\left[\frac{\gamma_1}{8\pi \alpha^2}\frac{|\alpha|}{\beta_1}\right]^{1/2}.
\end{eqnarray}
One measures lengths in units of $l_{m}$: $R=r/l_m $. We also introduce
\begin{equation}
h_0 = \frac{\phi_0}{\sqrt{8}\pi l_{m}l_{\xi}}, ~~ \mathbf{A}(\mathbf{r}) = \sqrt{2}l_{m} h_0 \mathbf{a}(\mathbf{R}),~~ \phi_0 = \frac{\pi c}{|e|}
\end{equation}
such that magnetic field $\mathbf{B} =\boldsymbol{\nabla}_R\times\mathbf{a}$ is measured in units of $h_0$, together with the following rewriting 
\begin{eqnarray}
\mathbf{D}= -i\bnabla_r -\frac{2e}{c}\mathbf{A}(\mathbf{r}) \rightarrow \mathbf{D} = -\frac{i}{\kappa}\bnabla_R +\mathbf{a}(\mathbf{R}),
\end{eqnarray}
where 
\begin{eqnarray}\label{Eq:kappa}
\kappa = l_m/l_{\xi}.
\end{eqnarray}
 We obtain
\begin{eqnarray}\nonumber
\mathcal{F} &=&\frac{h^2_0 l^3_m}{4\pi}\sum_{s=\pm} \int d^3R\bigg\{ -|\psi_{s}|^2+ (D_x\psi_{s})^*D_x\psi_{s}+(D_y\psi_{s})^*D_y\psi_{s}+\beta_z (D_z\psi_{s})^*D_z\psi_{s}
\\
&+&
\beta_{\perp}(D_{-s}\psi_{s})^*D_{s}\psi_{-s}+\frac{|\psi_{s}|^4}{2}+
\frac{\gamma}{2}|\psi_{s}|^2|\psi_{-s}|^2\bigg\}
\end{eqnarray}
where we introduce new coefficients:
\begin{equation}
\beta_{\perp} = \frac{\beta_2}{\beta_1},~~ \beta_{z} = \frac{\beta_3}{\beta_1},~~ \gamma=1-\frac{2\gamma_2}{\gamma_1}.
\end{equation}
They are not independent. First is the symmetric case where we set $v=v_z$ and obtain
\begin{equation}
\beta_{\perp} = \frac{1}{3}, \beta_{z}=\frac{4}{3}, \gamma=\frac{1}{3}<1
\end{equation}
treating $\kappa$ as the only free parameter. Second, is the cylindrical case where we set $v_z=0$ and obtain
\begin{equation}
\beta_{\perp} = \frac{1}{2}, \beta_{z}=0, \gamma=2>1
\end{equation} 
again treating $\kappa$ as the only free parameter. Finally, including Maxwell term $(\bnabla_R\times \mathbf{a})^2$ into free energy one obtains

\begin{eqnarray}
\mathcal{F}_t &=&\frac{h^2_0 l^3_m}{4\pi}\sum_{s=\pm} \int d^3R\bigg\{ -|\psi_{s}|^2+ |D_x\psi_{s}|^2+|D_y\psi_{s}|^2+\beta_z |D_z\psi_{s}|^2
\\
&+&\nonumber
\beta_{\perp}(D_{-s}\psi_{s})^*D_{s}\psi_{-s}+\frac{|\psi_{s}|^4}{2}+
\frac{\gamma}{2}|\psi_{s}|^2|\psi_{-s}|^2 + (\bnabla_R \times \mathbf{a})^2\bigg\}
\end{eqnarray}

\end{document}